\documentclass[aps,prl,reprint,twocolumn,superscriptaddress,showpacs]{revtex4-1}

\usepackage{amsmath}
\usepackage{graphicx}
\usepackage{amsfonts}
\usepackage{color}
\usepackage[usenames,dvipsnames]{xcolor}
\usepackage[caption=false]{subfig}

\newcommand{\ssc}[0]{\sin{\gamma}}
\newcommand{\ccc}[0]{\cos{\gamma}}

\newcommand{\cct}[0]{\cos{\theta}}
\newcommand{\ssp}[0]{\sin{\phi}}
\newcommand{\ccp}[0]{\cos{\phi}}
\newcommand{\vcrm}[1]{\mathbf{#1}}
\newcommand{\hvcrm}[1]{\mathbf{\hat{#1}}}
\newcommand{\vc}[1]{\boldsymbol{#1}}

\newcommand{\dd}{\mathrm{d}}
\newcommand{\ee}{\mathrm{e}}

\newcommand{\kk}{\mathrm{k}_B}
\newcommand{\vm}{\vc{\mu}}
\newcommand{\vn}{\hvcrm{n}}

\begin{document}

\title{Thermal analogue of gimbal lock in a colloidal ferromagnetic Janus rod}

\author{Yongxiang Gao}
\email{yongxiang.gao@chem.ox.ac.uk}
\affiliation{Department of Chemistry, Physical and Theoretical Chemistry Laboratory, University of Oxford, South Parks Road, Oxford OX1 3QZ, United Kingdom}
\author{Andrew Kaan Balin}
\email{andrew.balin@physics.ox.ac.uk}
\affiliation{Rudolf Peierls Centre for Theoretical Physics, 1 Keble Road, University of Oxford, OX1 3NP, United Kingdom}
\author{Roel P.A.\ Dullens}
\affiliation{Department of Chemistry, Physical and Theoretical Chemistry Laboratory, University of Oxford, South Parks Road, Oxford OX1 3QZ, United Kingdom}
\author{Julia M.\ Yeomans}
\affiliation{Rudolf Peierls Centre for Theoretical Physics, 1 Keble Road, University of Oxford, OX1 3NP, United Kingdom}\author{Dirk G.A.L.\ Aarts}
\affiliation{Department of Chemistry, Physical and Theoretical Chemistry Laboratory, University of Oxford, South Parks Road, Oxford OX1 3QZ, United Kingdom}

\date{\today}

\begin{abstract} We report an entropy-driven orientational hopping transition in a magnetically confined colloidal Janus rod. In a magnetic field, the sedimented rod randomly hops between horizontal and vertical states: the latter state comes at a substantial gravitational cost at no reduction of magnetic potential energy. The probability distribution over the angles of the rod shows that the presence of an external magnetic field leads to the emergence of a metastable vertical state separated from the ground state by an effective barrier. This barrier does not come from the potential energy but rather from the vast gain in phase space available to the rod as it approaches the vertical state. The loss of rotational degree of freedom that gives rise to this effect is a statistical mechanical analogue of the phenomenon of \emph{gimbal lock} from classical mechanics.\end{abstract}

\pacs{05.20.Gg, 47.7.J, 45.20.dc, 65.80.-g}

\maketitle

In soft matter, many phenomena are governed by entropy. Examples include polymer translocation through microchannels \cite{Muthukumar1989,Ledesma-Aguilar2012,Zhang2012}, accumulation of colloidal spheres at corners of a rough substrate \cite{Dinsmore1996} and particle diffusion through constrictive environments or around obstacles \cite{Chou1999,Zwanzig1992,Ding2015}. Boundary effects and confinement play a key role in many such instances and it is likely that biological processes exploit many of these phenomena between the molecular and cellular levels. Non-spherical colloids in particular give rise to very rich behavior in different regimes. For instance, depletion-induced torque on rigid rods near a wall have been suggested to play a role in the biochemical key-lock mechanism \cite{Roth2002,Helden2003} while in the very high rod-density limit, the statistical mechanics of nematic liquid crystals is currently of great importance to the study of active biological matter \cite{Baskaran2009,Marchetti2013}.

There has been growing interest in examining the rotationally driven dynamics of individual rod-like colloids \cite{Dhar2007,Shelton2005,Ghosh2012,Rikken2014}, especially for their potential applications as microtools \cite{Solovev2012,Xi2013}. The rod-like or helical particles considered in these studies possess no continuous rotational symmetry in their internal structure, i.e.\ continuous rotation about any axis will alter their energy. In this Letter, we demonstrate and analyse the entropic hopping behavior of a ferromagnetic rod under the influence of a static external magnetic field. The rod is observed to transition between horizontal and vertical orientations with respect to the substrate, despite the latter state corresponding to an overall gain of $\sim3\ \kk T$ in potential energy. Its long-time distribution of polar angles reveals two peaks (corresponding to vertical and horizontal states) separated by an effective barrier. This system is simple enough to yield an accurate analytical description with no adjustable parameters, revealing the entropic nature of this behaviour.

We grew silica particles in the shape of hemispherically capped cylinders, each with total length $L$ between $3$-$4  \mu$m and diameter $d$ ranging $0.6$-$0.7 \mu$m \cite{Gao15}. The synthesis protocol is based on a modification of that presented in Ref.\ \cite{Kuijk2011}. Each rod was doped with magnetite (Fe$_3$O$_4$) nanoparticles in one of the hemispherical caps, hence we call them Janus rods. After the application of a saturating external magnetic field, the caps retain a net magnetization $\vc{\mu}$ in the cross-sectional plane of each rod, perpendicular to the rods' long-axis. These particles were then suspended in deionized water (Millipore, 18.2 M$\Omega$) at a very low volume fraction ($\approx 10^{-6}$) and loaded into a custom glass cell (inner dimension $2\times0.5\times0.15$ cm$^3$). The sample was allowed to rest on a microscope stage for 10 minutes until the particles formed a sediment. A magnetic field was applied by a pair of Helmholtz coils with an approximate range of $0$-$150$ gauss (G). All experiments were conducted at room temperature on an inverted light microscope equipped with a $60\times$ oil-immersion lens (NA=1.42). Bright field images were acquired at 20 frames per second.

\begin{figure}
		\includegraphics[width=0.95\columnwidth]{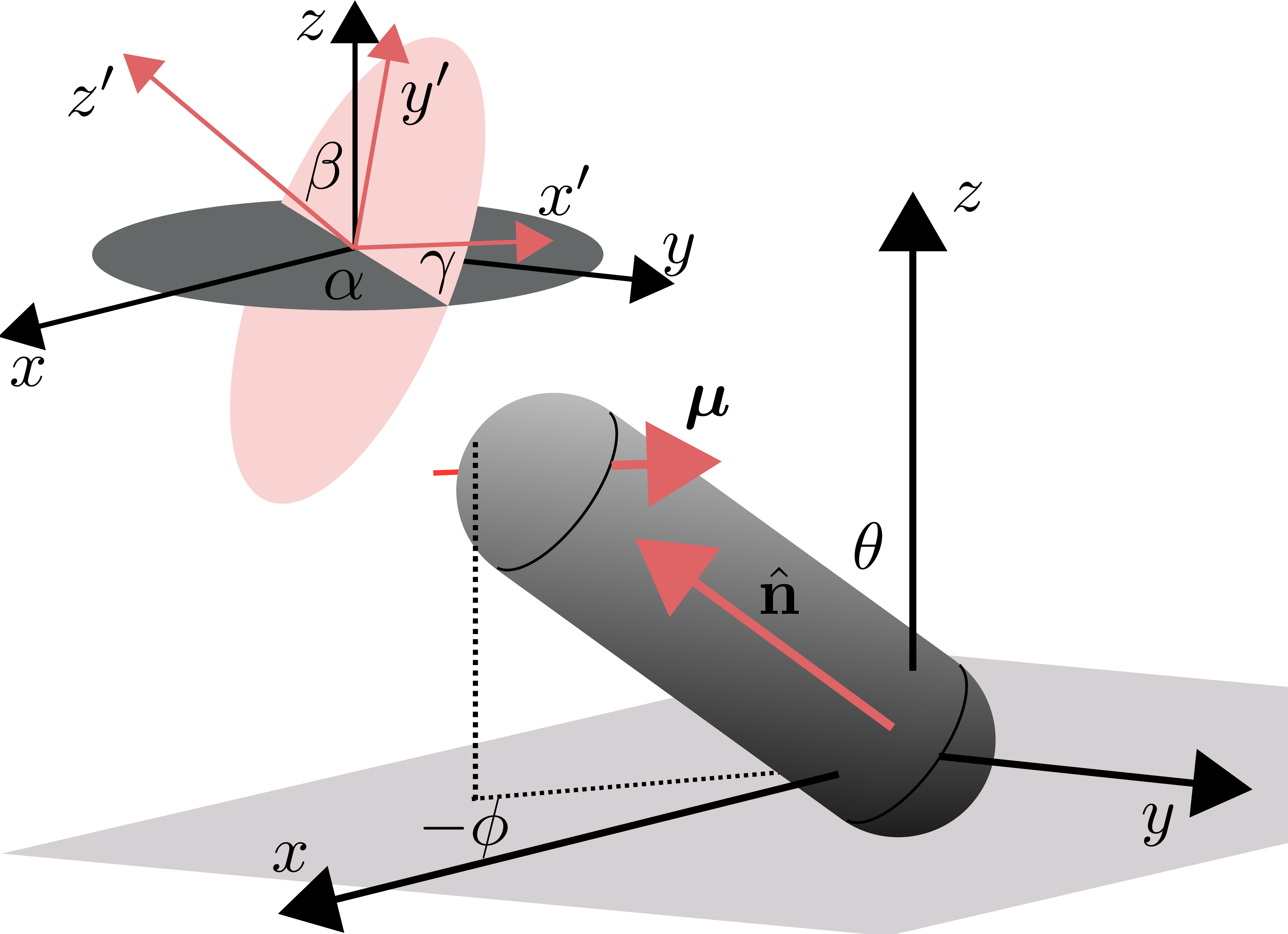}
	\caption{\footnotesize (Color Online) \emph{Main:} Geometry of an arbitrarily oriented rod. The rod's long axis is spanned by $\hvcrm{n}$ and makes an angle $\theta$ with the $z$-axis. Its projection in the $xy$-plane subtends an angle $\phi$ with the $x$-axis. A permanent magnetic moment $\vc{\mu}$ is embedded in one end of the rod and rotates rigidly with it. The gravitational field acts in the $-$ve $z$-direction, and an external magnetic field $\vcrm{B}$, when applied, is along the $x$-direction, causing the rod to rotate from the configuration shown to one where $\vc{\mu}$ aligns with $\vcrm{B}$. \emph{Inset:} Euler angles for describing the fixed-body rotation of the rod, where $\hvcrm{n}=\hvcrm{z}'$, $\vc{\mu}=\mu\hvcrm{x}'$, $\beta=\theta$, and $\alpha=\phi+\frac{\pi}{2}$.\label{fig:geometry}}
\end{figure}

These rods are similar in geometry (shown in Fig.\ \ref{fig:geometry}) to those studied in the dynamical rotating-field experiments by Dhar \emph{et al.}\ \cite{Dhar2007}. To describe the long-axis of the rod we use a spanning vector $\hvcrm{n}$ which in the laboratory frame makes an angle $\theta$ with the $z$-axis, and an angle $\phi$ in the $xy$-plane. One cap of the rod has an embedded permanent magnetic moment $\vc{\mu}$ that is perpendicular to $\hvcrm{n}$ and requires a third angle $\gamma$ to parameterise its direction in the cross-sectional plane of the rod. If we consider the fixed-body coordinates of the rod, $(x',y',z')$, then $\hvcrm{n}$ lies along the $z'$-axis and $\vc{\mu}$ lies in the $x'z'$-plane which in general breaks axial symmetry. The conventional $Z_\alpha X_\beta Z_\gamma$ Euler angles $(\alpha=\phi+\pi/2, \beta=\theta, \gamma)$ describe the rotation of the rod-frame relative to the lab-frame. Hence, each state of the rod $(\vm,\vn)$ can be described in terms of these angles and gives rise to an instantaneous energy $U=-\vc{\mu}\cdot\vcrm{B} - m^*gh$, where $m^*=\Delta\rho V$ is the effective mass and $h$ is the height of the centre-of-mass of the rod above its minimum ($d/2$ when the rod lies flat).

Due to its large density relative to water, ($\Delta\rho = \rho_r - \rho_w \approx 0.9 \cdot 10^3$ kg m$^{-3}$ \cite{Kuijk2011}) a rod quickly sediments on the coverslip of the microscope slide. Assuming contact with the surface, $h=-\frac{1}{2}g(L-d)\hvcrm{n}\cdot\vc{z}$ allows us to write $U$ entirely as a function of ($\vc{\mu},\hvcrm{n}$):\begin{equation}\label{energy}
U(\vc{\mu},\hvcrm{n})= -\vc{\mu}\cdot\vcrm{B} + \frac{m^* g l}{2}  \hvcrm{n}\cdot\vc{z}
\end{equation}where $l=L-d$ is the length of the cylindrical section of the rod.

The rod lies horizontally in the plane ($\theta\lesssim\pi/2$) whilst undergoing rotational Brownian motion in $\phi$ when no external field is present. Thermal deviation far below $\theta=\pi/2$ is exponentially suppressed by gravity. In the presence of an in-plane magnetic field $\vcrm{B}=B \hvcrm{x}$, the magnetic moment $\vc{\mu}$ aligns with $\vcrm{B}$, trapping the rod in the $yz$-plane. The azimuthal angle of the rod fluctuates about either of the points $\phi_0=\pm\pi/2$. We make use of the equipartition theorem $ \frac{1}{2}\kk T = \langle -\vc{\mu}\cdot\vcrm{B}\rangle \approx \frac{1}{2}\mu B \langle \Delta\phi^2 \rangle $ to calculate the strength of the moment $\mu$ by measuring the fluctuations $\Delta\phi=\phi-\phi_0$ at varying field strengths at room temperature. The spread of $\Delta\phi$ decreases for larger $B$, and does so in a manner consistent with $\mu(B)$ remaining constant across the full range of fields applied, from which we infer that the rod cap is ferromagnetic. Typically, we measured the strength of the magnetic moment to be approximately $1$-$2\ \kk T $ G$^{-1}$ at room temperature.

The main experimental observation that motivated this study was that confinement of the rods to the $yz$-plane by a magnetic field resulted in the emergence of an apparent bistability between vertical ($\theta \approx 0$) and horizontal ($\theta \lesssim \pi/2$) orientations, with thermal fluctuations alone strong enough to explore both states ---in contrast to an energy-consuming excitation-relaxation process or a driven process resulting in similar behaviour \cite{Dhar2007}. Importantly, this effect occurs in spite of the fact that the rod gains $\frac{1}{2}m^* g l \approx 3.4 \ \kk T$ of gravitational potential energy at no reduction of magnetic energy $\vc{\mu}\cdot \vcrm{B}$ (as $\vc{\mu}$ is able to remain aligned with the field at all times).

\begin{figure}[t]
\centering
\includegraphics[width=\columnwidth]{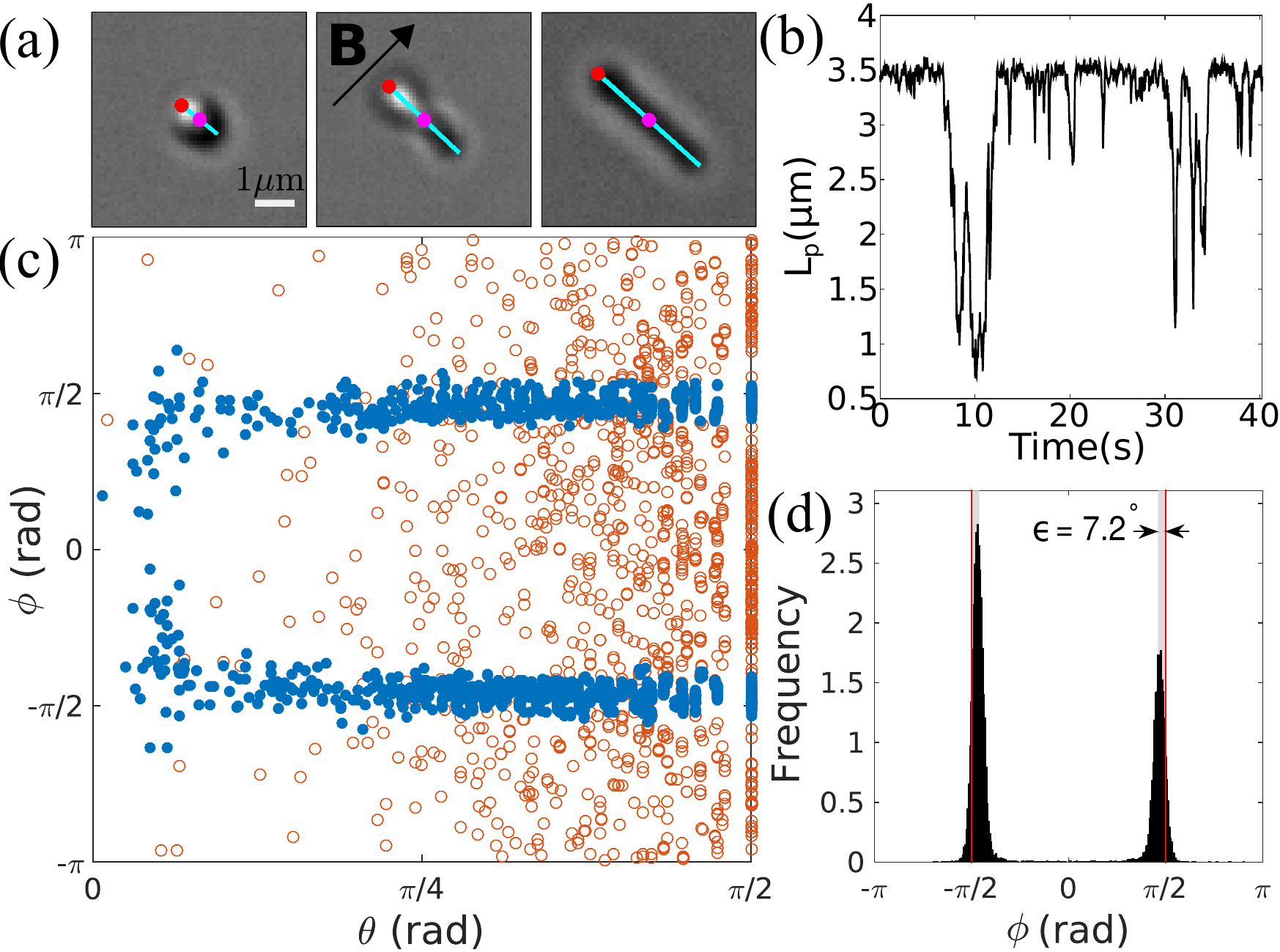}
\caption{\footnotesize \label{fig2} (Color online) (a) Typical optical images in $xy$-plane with a rod confined by a field. Overlaid are the centroid ({\color{magenta}$\bullet$}), length and orientation ({\color{cyan}--}) and the ferromagnetic cap ({\color{red}$\bullet$})of the rod. (b) A short segment of projected length $L_p(t)$ demonstrates the hopping behaviour.(c) Scattered [$\theta(t)$,$\phi(t)$] for no field ({\color{red}$\circ$}) and $140$ G field ({\color{NavyBlue}$\bullet$}) demonstrates the magnetic trapping (for clarity, we display every tenth sequential point). (d) Histogram of $\phi(t)$ for $B=140$ G case enables us to calculate the magnetic moment's offset angle $\epsilon$ by measuring the shift of the peak from $\pm \pi/2$.}
\end{figure}

This dynamic behavior is best described as a hopping transition between these two states occurring at random (though with different characteristic forwards- and backwards-rates) and is apparent in real-time video data \cite{SM1}. For illustration, three frames captured at different times are displayed in Fig.\ \ref{fig2}a. By automated analysis of the images \cite{SM2}, we measured the projected length $L_p(t)$ of the rod in the $xy$-plane at each time, a segment of which is shown in Fig.\ \ref{fig2}b and clearly demonstrates the hopping behaviour between the horizontal state (large $L_p$) and the vertical state (small $L_p$). We also measured the azimuthal angle $\phi(t)$ of the rod in the image plane. We calculated the polar angle using $\theta(t) = \sin^{-1}{[(L_p(t)- d)/(L - d)]}$, where $L=3.5\ (\pm0.1)\ \mu$m and $d=0.65\ (\pm0.03)\ \mu$m are the measured total length and diameter of the rod respectively. Due to the measurement error of these, occasionally, $L_p(t)$ falls out of the range $[d, L]$. In these cases, we set it to its corresponding bounding value in order for $\theta(t)$ to remain real-positive.

\begin{figure}
\centering
\subfloat[][]{\includegraphics[width=0.48\columnwidth]{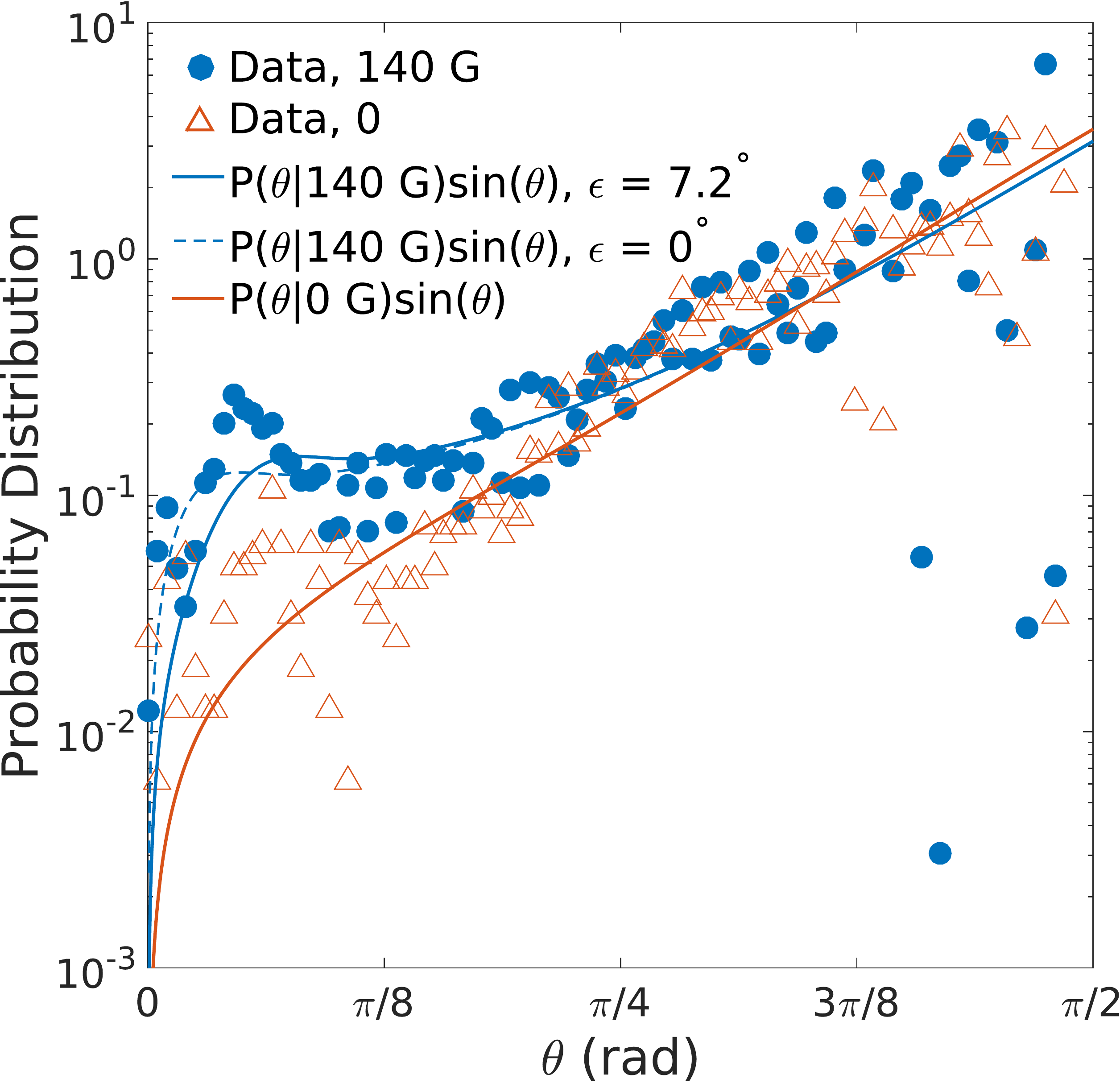}\label{Pdata}}
\subfloat[][]{\includegraphics[width=0.48\columnwidth]{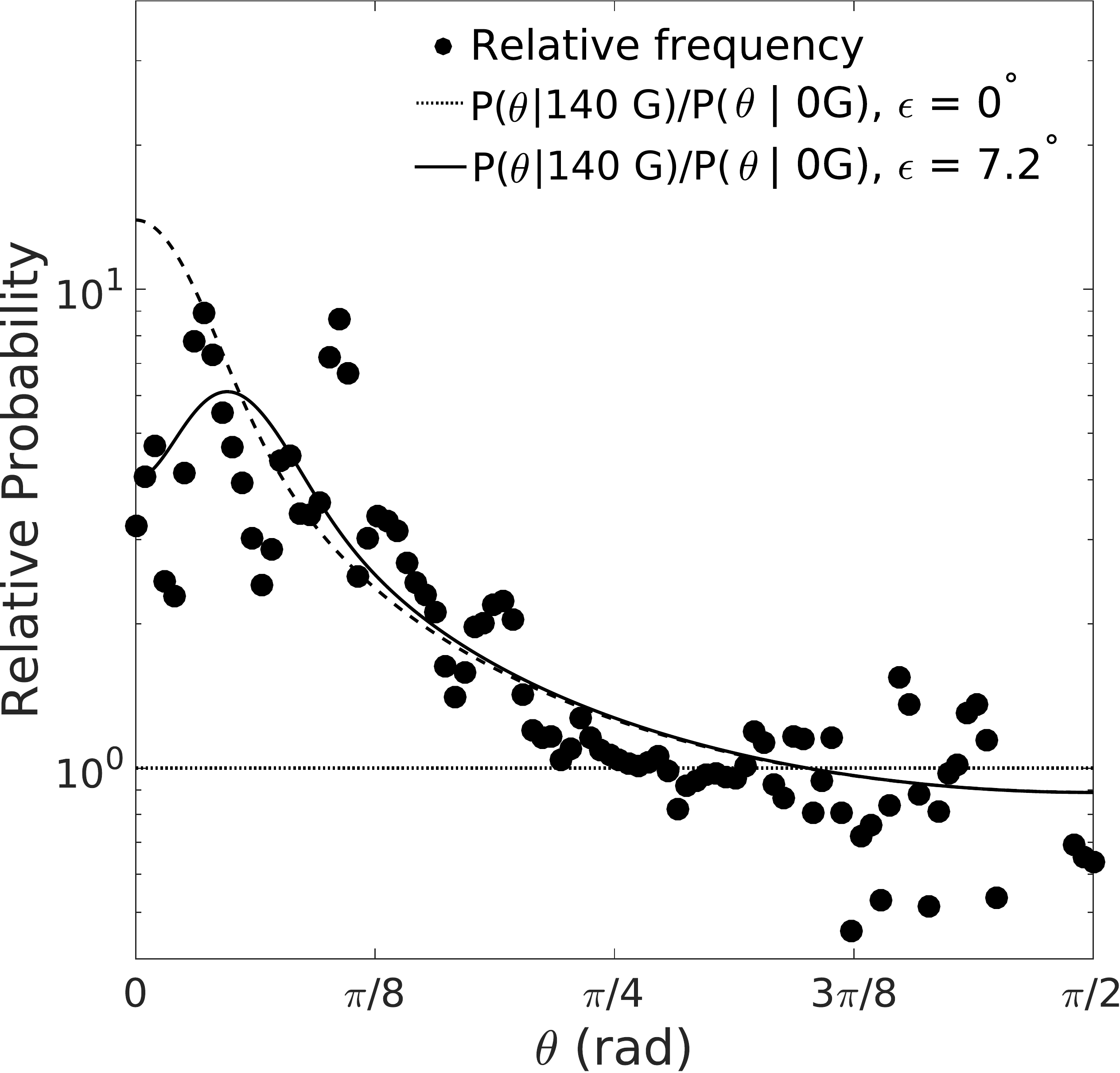}\label{ratio}}
    \caption{\footnotesize (Color Online) (a) Histogram of $\theta(t)$ shows a markedly increased tendency for vertical states (small $\theta$) to be realised when a magnetic field is present ({\color{NavyBlue}$\bullet$}) with respect to a free rod ({\color{red}$\bigtriangleup$}). The curves represent absolute probability weightings $P(\theta) \sin(\theta)$ where we have numerically integrated Eq.\ (\ref{corrected}) using independently measured parameters $\mu$, $\epsilon$, $m$, $L$ and $d$. (b) The relative distribution $P(\theta | 140 $G$)/P(\theta | 0 $G$)$ agrees well with the theoretical calculations, showing an $O(10)$ increase in likelihood for small-$\theta$ states due to the magnetic field compared to no field.}
\end{figure}

To demonstrate that the presence of a magnetic field is primarily responsible for the hopping behaviour, we ran two experiments for 10 and 20 minutes each at $0$ G and $140$ G field strengths respectively. For these experiments, we used a rod with a moment strength measured to be $\mu = 1.1\pm0.1\ \kk T$ G$^{-1}$. Figure \ref{fig2}c qualitatively demonstrates both the trapping in $\phi$ and the increase in instances of low-$\theta$ measurements in the magnetically confined case. However Fig.\ \ref{fig2}d exposes a more subtle feature of the data: rather than being confined exactly along the $y$-axis ($\phi_0=\pm\pi/2$), the mean azimuthal angle of the rod deviated from this by around $\epsilon = (7.2\pm 0.1)^\circ$. This is evidence that rather than being precisely within the cross-sectional plane of the rod, $\vc{\mu}$ makes an angle $\epsilon\approx 7.2^\circ$ with it. In the next section, we will theoretically consider both the ideal case $\vc{\mu}\cdot\hvcrm{n}=0$, as well as an exact description accounting for finite $\epsilon$. Figure \ref{Pdata} contains histograms of $\theta(t)$ both with and without the field and the histogram ratio (Fig.\ \ref{ratio}) clearly shows how the magnetic field accounts for an $\sim O(10)$ increase in relative likelihood of the rod being found in the vertical state compared to the no field case. Rather than having a peak at exactly $\theta=0$, due to the finite $\epsilon$, the maximum-likelihood vertical state is offset by about $\sim 0.15$ rad.

The theoretical curves plotted on top of the data in Figs.\ \ref{Pdata}-\ref{ratio} are calculated by assuming that on times much longer than the dynamical timescales, the states ($\vc{\mu},\hvcrm{n}$) form a canonical ensemble with energies given by Eq.\ (\ref{energy}). As we want to end up with a distribution over $\theta$, we evaluate each term in the rod's fixed body frame $(x',y',z')$, where $\vc{\mu}' :=\mu(\cos\epsilon,0,\sin\epsilon)$ and $\hvcrm{n}' :=(0,0,1)$. For an ideal rod, $\epsilon=0$ means the magnetic moment is orthogonal to the axis of the rod. The external fields in this frame are thus given by solid body rotations: $\vcrm{B}' = \vcrm{R}(\phi,\theta,\gamma)\cdot ( B \hvcrm{x})$, and $\vcrm{g}'= \vcrm{R}(\phi,\theta,\gamma)\cdot ( -g \hvcrm{z})$, where $\vcrm{R}(\phi,\theta,\gamma)$ is the rotation matrix describing the laboratory- to rod-frame transformation. It can be shown that\begin{eqnarray}
U(\phi,\theta,\gamma)  = & - & \mu B \Big( \cos\epsilon [\ccc\ssp + \cct\ccp\ssc] \nonumber \\
& + & \sin\epsilon \cos\phi\sin\theta \Big) + \frac{m^*gl}{2} \cct.
\end{eqnarray}
The equilibrium distribution of the rod is the Boltzmann distribution $P(\theta, \phi, \gamma)  =  \frac{1}{Z} \ee^{ -U(\phi, \theta, \gamma)/\kk T }$ where $Z$ is the partition function.

As $\gamma$ is not measured by experiment we wish to first find the marginal distribution $P(\theta,\phi)$ by integrating out the $\gamma$ dependence. To simplify the notation, we introduce the relative strength of the gravitational energy $a=m^*gl/2 \kk T$ and magnetic energy $b=\mu B/\kk T$. The integral may be carried out explicitly, giving
\begin{eqnarray}\label{corrected}
P(\theta,\phi)  & = & \frac{\ee^{-a\cos\theta}}{Z} \Big\{ I_0\Big( b\cos\epsilon\sqrt{1-\cos^2\phi\sin^2\theta} \Big) \nonumber\\
& \times &  \ee^{-b \sin\epsilon \cos\phi\sin\theta}\Big\}, \label{P_corrected}
\end{eqnarray}
where $I_0(x)$ is the modified Bessel function of the first kind. This distribution contains no adjustable parameters as $a=3.4$, $b=156$, and $\epsilon=7.2^\circ$ are all independently measurable for a given rod and magnetic field.

We numerically integrate Eq.\ (\ref{P_corrected}) with respect to $\phi$, to obtain the marginal distribution $P(\theta)$. To compare the $P(\theta)$ with experiment, we plot the absolute probability $P(\theta)\sin(\theta)$ on top of the histogram of $\theta(t)$ in Fig.\ \ref{Pdata}. Both are in good agreement and the $140$ G data shows a clear peak at $\theta=0.15$ rad. There appears to be a slight systematic excess in the low-$\theta$ region of the data. This is likely attributable to the image analysis procedure uniformly underestimating $L_p$ for low $\theta$. However this error is factored out when we plot the relative likelihood $P(\theta | 140$ G$)/P(\theta | 0$ G$)$ on top of the corresponding ratio of the histograms (Fig.\ \ref{ratio}) which show the differential effect of exposing the rod to a magnetic field. 

\begin{figure}
\centering
\subfloat[][]{\includegraphics[width=0.5\columnwidth]{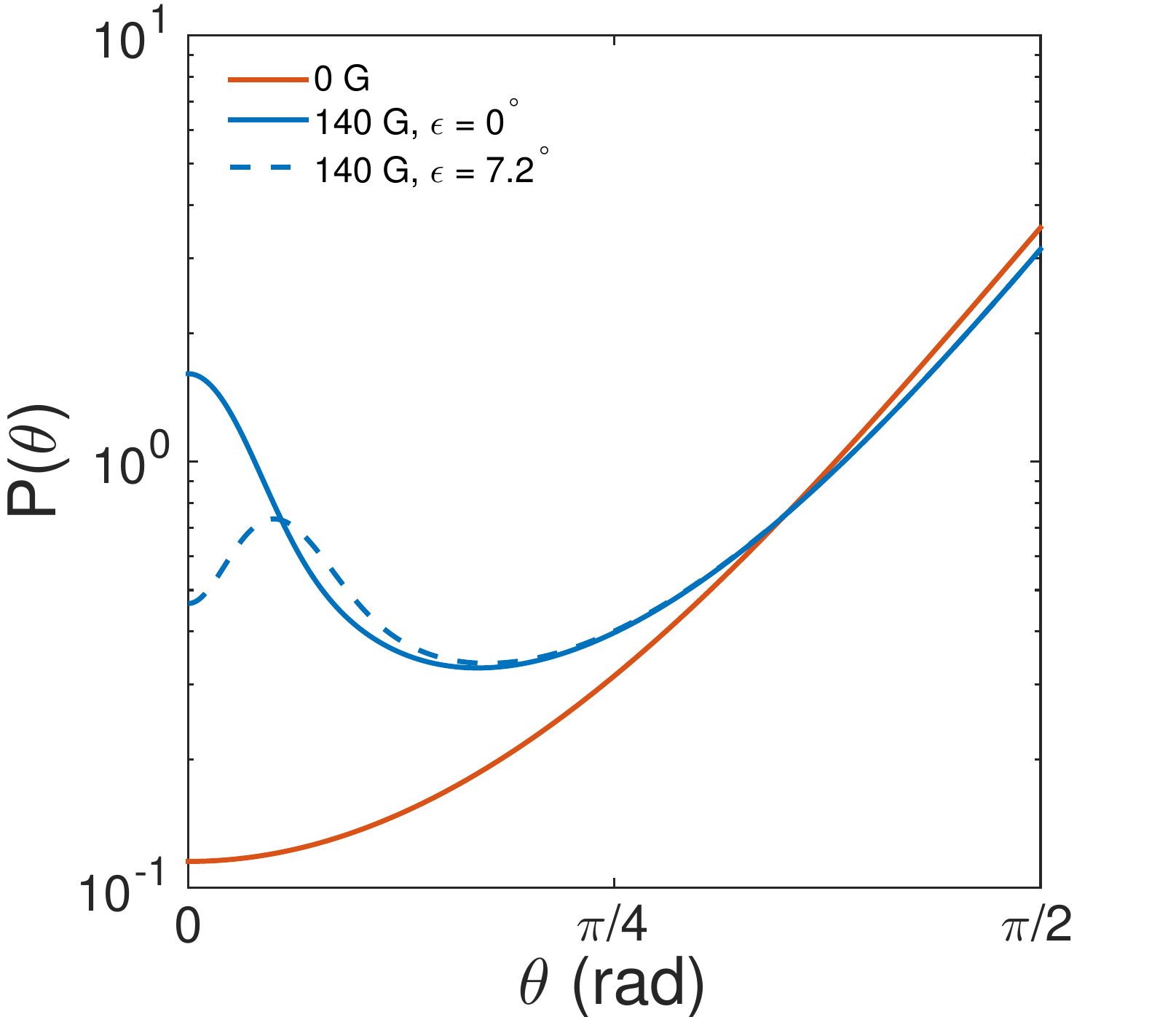}\label{theorya}}
\subfloat[][]{\includegraphics[width=0.5\columnwidth]{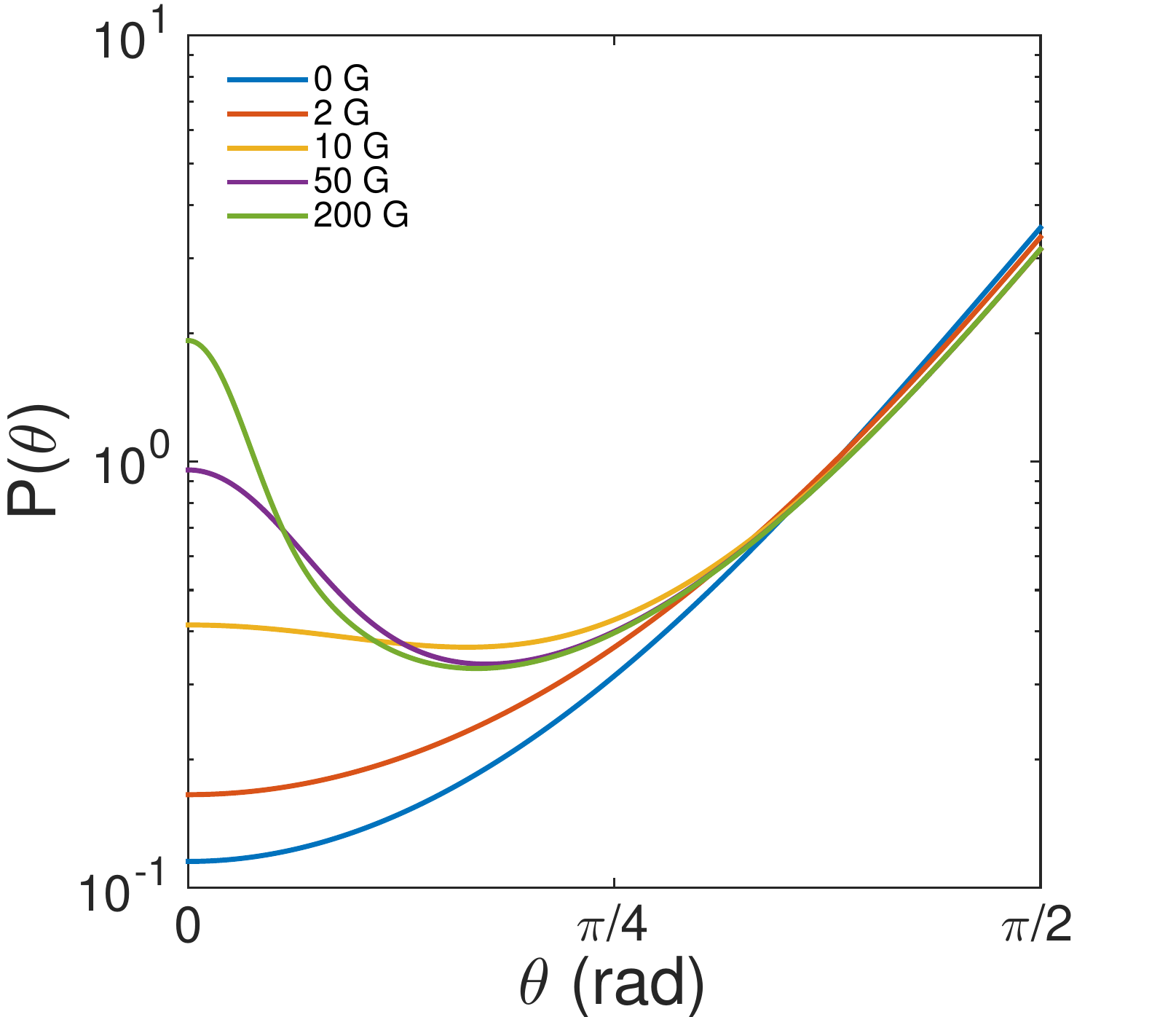}\label{theoryb}} \\
\subfloat[][]{\includegraphics[width=\columnwidth]{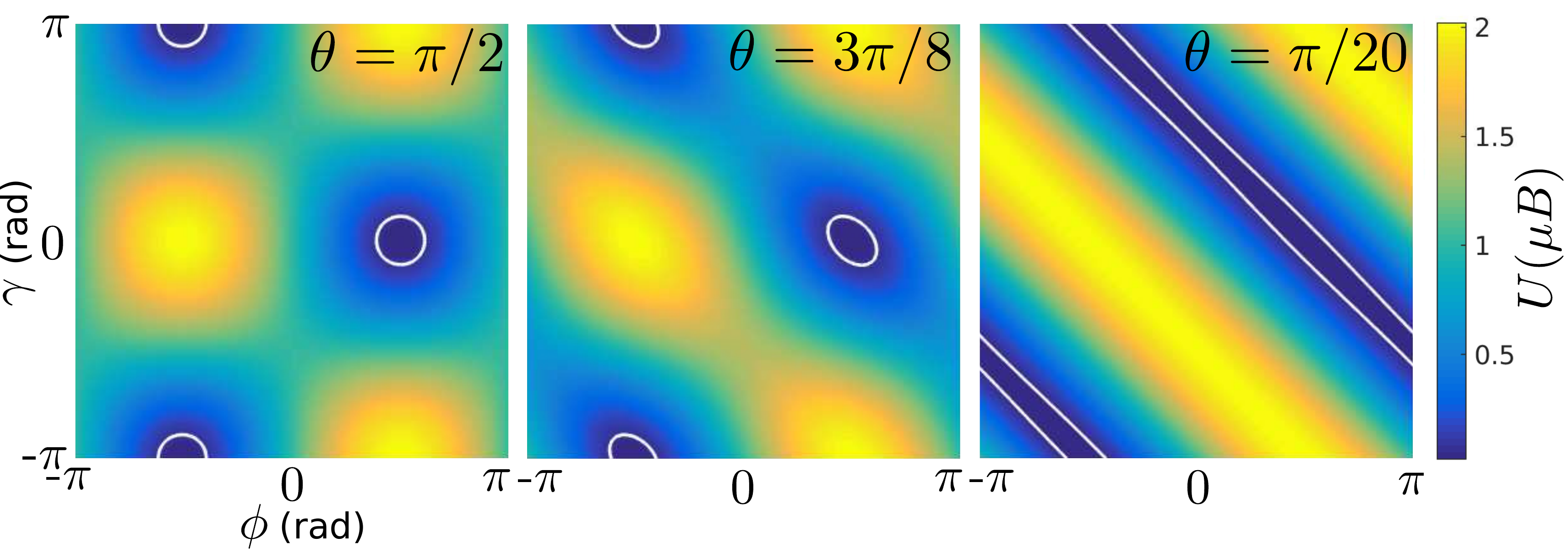}\label{phase}}
    \caption{\footnotesize (Color Online) (a) Probability density function for a rod with $\mu=1.1\ \kk T$ G$^{-1}$ in $140$ G (blue) and $0$ G (red) fields shows an emergent bistability between vertical and horizontal states due to the external field for an ideal rod (---) and one where $\vc{\mu}$ is offset from the cross-sectional plane by $\epsilon=7.2^\circ$ (- -). (b) Same as (a) but for varying magnetic fields for an ideal rod. (c) The energy $U(\phi,\theta=\Theta,\gamma)$ for three polar angles $\Theta=\frac{\pi}{2},\frac{3\pi}{8},\frac{\pi}{20}$ shows the loss of a degree of freedom. When the rod is lying flat, $\phi$ and $\gamma$ are independently constrained, but when vertical only the compound angle $\phi+\gamma$ is constrained. This results in the entropic favouring of vertical states compared to intermediate states despite the gravitational cost. We have plotted the energy relative to the minimum energy $U_{0}=-\mu B$.\label{theory}}
\end{figure}

The marginal density $P(\theta)$ on its own (Figs.\ \ref{theorya}-\ref{theoryb}) for both the experimental system (with $\epsilon=7.2^\circ$) and that for an ideal rod ($\epsilon=0^\circ$) under varying magnetic fields shows clearly the transition from monostability to bistability. This transition occurs at around $b > O(10)$, and becomes strong (i.e.\ $P(\pi/2)/P(0) \approx O(1)$) at around $b=O(100)$. The overall effect of applying a large field is to encourage the population of low-$\theta$ states that are otherwise gravitationally suppressed, revealing a minimum at $\theta_{\text{min}}\ne 0$. Considering the quantity -$\ln P(\theta)$, we interpret there as being an effective barrier separating two bistable states. We identify this as an entropic barrier because it does not come from the potential energy which is a monotonic function of $\theta$ (keeping $\vc{\mu}\cdot\vcrm{B}$ minimised with the angles $\phi=\pi/2,\gamma=0$ confined by the field).

We wish to interpret the marginal distribution $P(\theta)$. If we consider a generalised potential $U(q_1,q_2)$ then in units where $\kk T=1$ we have
\begin{equation}\label{log-marginal}
\ln{P(q_1)} = \ln\Big[ \int_{Q_2} \dd q_2\ \ee^{-U(q_1,q_2)} \Big] + F,
\end{equation}with $F=-\ln Z$. If $q_1$ and $q_2$ are independent (as in most familiar cases, e.g.\ harmonic potentials $U(x,y)= k_x x^2 + k_y y^2$) then $\ln P(q_1)$ is proportional to the potential energy landscape $U_1(q_1)$ up to an additive constant. However when this is not the case ---for instance in cases like ours where the coordinates are coupled by the potential--- then Eq.\ (\ref{log-marginal}) is not so readily interpreted as a thermodynamic quantity. However, if we take the negative partial derivative with respect to $q_1$, we get
\begin{eqnarray}
-\frac{\partial }{\partial q_1} \ln{P(q_1)} & = & \frac{\int_{Q_2} \dd q_2 \ \frac{\partial U(q_1,q_2)}{\partial q_1} \ee^{-U(q_1,q_2)} }{\int_{Q_2} \dd q_2 \  \ee^{-U(q_1,q_2)}}\nonumber \\ 
& = & \Big\langle \frac{\partial U(q_1,q_2)}{\partial q_1} \Big\rangle_{q_2}.
\end{eqnarray}
The right-hand-side of this has the form of an effective generalised force which from the form of the left-hand-side we identify as an entropic force \cite{Neumann1980,Roos2014}. For a rod at an angle $\theta=\Theta$, this is an effective torque
\begin{equation} \label{entropicforce}
-\frac{\partial }{\partial \theta}\Big|_{\theta=\Theta} \ln{P(\theta)}  = \Big\langle \frac{\partial U(\theta,\phi,\gamma)}{\partial \theta}\Big|_{\theta=\Theta} \Big\rangle_{\phi,\gamma},
\end{equation}where the average is taken over the angles $\phi,\gamma$.

Naively, one would think that for a rod that has found its global energy minimum where $\vc{\mu}$ is aligned along $\vcrm{B}$ while lying flat, the fact that any change in $\theta$ acts only to increase the gravitational potential energy ---while leaving $\vc{\mu}\cdot\vcrm{B} $ unchanged--- means such a change should be suppressed exponentially. However, Eq.\ (\ref{entropicforce}) tells us that we must take into account the cost of thermal excursions away from alignment of $\vc{\mu}$ with $\vcrm{B}$. When the rod is lying flat, $U\sim \cos\gamma\sin\phi$ means $\phi$ and $\gamma$ are independently and tightly constrained. In the opposite limit $\theta\rightarrow0$, the energy reduces to $U\sim \sin(\gamma+\phi)$ meaning only the compound angle $\phi+\gamma$ is constrained, opening up a much larger configuration-space available to the rod. This is demonstrated in Fig.\ \ref{phase} which shows how an iso-potential bounding low energy states deforms as $\theta$ is reduced. Between $\theta=\pi/2\rightarrow 3\pi/8$, the gravitational energy increases substantially so $P(\theta)$ decays as $\theta$ is decreased. However, between $\theta=3\pi/8\rightarrow\pi/20$ there is a smaller yet nevertheless positive gravitational cost, but the coinciding gain in available phase-space is large enough to compensate for this, hence the hopping that we observed experimentally is a true entropy-driven process. In classical mechanics, this loss of one degree of freedom is associated with gimbal lock: a phenomenon where a mechanical instrument controlled by Euler-like rotations irreversibly loses a degree of freedom when two axes coincide. Unlike this however, the rod in our system is not controlled by Euler-angle rotations; rather it is controlled by frame independent external forces which result from a potential energy which suffers a loss of degree of freedom when the rod aligns perpendicular to the substrate plane. While in mechanical systems, gimbal lock results in a reduction of control of the system, for a thermal system, an analogue of gimbal lock in the potential energy function results in an entropic gain of explorable phase space.

This effect highlights some subtle but fundamental physics. The broken axial symmetry due to the perpendicular moment means that the rod's potential $U$ suffers the inevitable loss of a degree of freedom associated with gimbal lock. This loss results in a reduction in the degree of confinement of the rod and so the vertical state becomes entropically favourable despite coming at a necessary and substantial energy cost. We have measured this for a colloidal rod, but indeed any particle with an energy possessing no orientational continuous symmetry may be susceptible to this kind of strong entropic effect in the correct temperature regime.

\begin{acknowledgments}

Y.G.\ and A.K.B.\ contributed equally to this work. Y.G. acknowledges support from Marie Curie actions (FP7-PEOPLE-2012-IIF No.\ 327919); J.M.Y. from an ERC Advanced Grant (291234 MiCE).

\end{acknowledgments}

\bibliography{refs.bib}
\end{document}